\documentclass[titlepage,aps,prd,twocolumn,groupedaddress,floatfix,nofootinbib,showkeys,11pt]{revtex4-2}

\usepackage[utf8x]{inputenc}
\usepackage{graphicx}
\usepackage{xurl}
\usepackage{setspace}

\usepackage{xcolor}

\begin{document}

\title{Fingerprinting Defects in Hexagonal Boron Nitride via Multi-Phonon Excitation}
\author
{Pablo Tieben,$^{1,2\ast}$ Andreas W. Schell,$^{1,2,3}$\\
$^{1}$Physikalisch-Technische Bundesanstalt, Bundesallee 100, 38116 Braunschweig, Germany\\
$^{2}$Institute for Solid State Physics, Gottfried Wilhelm Leibniz University,\\ 
Appelstrasse 2, 30167 Hannover, Germany\\
$^{3}$Institute of Semiconductor and Solid State Physics, Johannes Kepler University,\\
Altenberger Strasse 69, 4040 Linz, Austria\\
$^\ast$E-mail:  tieben@qute.uni-hannover.de
}


\begin{abstract}
Single photon emitters in hexagonal boron nitride have gathered a lot of attention due to their favourable emission properties and the manifold of possible applications. Despite extensive scientific effort, the exact atomic origin of these emitters has remained unkown thus far. Recently, several studies have tied the emission in the yellow spectral region to carbon-related defects, but the exact atomic structure of the defects remains elusive. In this study, photoluminescence emission and excitation spectroscopy is performed on a large number of emitters within this region. By comparison of the experimental data with theoretical predictions, the origin of yellow single photon emission in hexagonal boron nitride is determined. Knowledge of this atomic structure and its optical properties is crucial for the reliable implementation of these emitters in quantum technologies.
\end{abstract}

\maketitle

\begin{singlespace}

\section{Introduction}
Optical quantum technologies rely on the highly controlled generation of photonic quantum states. A promising way to generate such states in a scalable way are solid-state single photon emitters such as quantum dots or defect centers in diamond \cite{Review:solid-stateEmitters}. Each of these emitters comes with its specific advantages and disadvantages for usage in quantum technologies and there is an ongoing effort to discover new emitters with more favorable properties.
%%%%%%%%%%%%%%%%%%%%%%%%%%%%%%%%%%%%%%%%
\begin{figure*}[t]
\centering
\includegraphics[width=\textwidth]{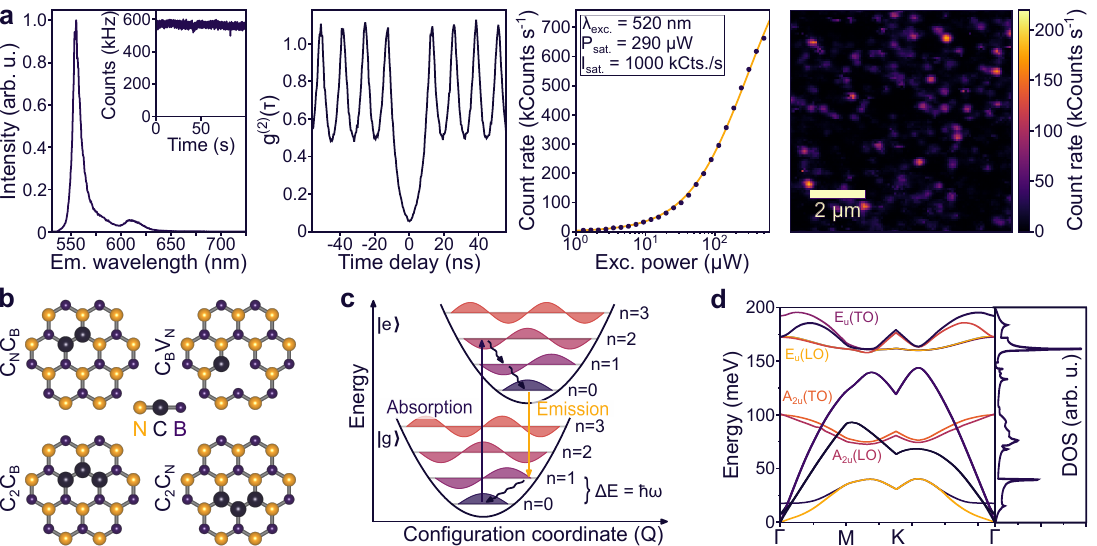}
\caption{Overview of an exemplary emitter characterization and the theoretical models used within this study. a) Emitter characterization. From left to right: Fluorescence emission spectrum with count rate trace with a binwidth of $10$\,ms (inset), autocorrelation, and saturation of an exemplary emitter and $8$\,x\,$8$\,µm\textsuperscript{2} confocal scan on the sample. All measurements are performed under $520$\,nm pulsed excitation with a laser power of $300$\,µW.
b) Four different proposed carbon defects $C_NC_B$ ($C_2$) (top left), $C_BV_N$ (top right), $C_2C_B$ (bottom left) and $C_2C_N$ (bottom right).
c) Huang-Rhys model showing vibronic states of the ground and excited state as well as exemplary phonon assisted transitions responisble for phonon sidebands and phonon assisted excitation.
d) Ab-initio calculation of the phonon dispersion and respective density of states for bulk hBN (after \protect\cite{LayersDependent}).}
\label{fig:1}
\end{figure*}
%%%%%%%%%%%%%%%%%%%%%%%%%%%%%%%%%%%%%%%%
After the first report in 2016, color centers in hexagonal boron nitride (hBN) have gained increased attention \cite{FirstReport}. Being hosted in a layered two-dimensional material, they are fundamentally different from defects in bulk materials such as diamond. Their remarkable optical properties together with their quantum emission at room-temperature make these emitters excellent candidates for technological aplications like integrated photonic structures \cite{TaperedFiber,Plasmonic1,Plasmonic2,StrainEngineering,StrainTuning} and quantum cryptography \cite{QKD1Edited,QKD2}.
Defects in hBN exhibit bright and stable emission with high single photon purity at and above room temperature \cite{800K} with zero phonon lines (ZPLs) ranging over a broad spectrum from the near-ultraviolet to near-infrared \cite{UV,UVtoNIR,BlueEmitters,UltraBrightMultiColor,Review:hBNAReview,NIR}. Certain defects have been shown to possess Fourier-limited linewidth at room temperature \cite{MechanicalDecoupling}, tunable internal quantum efficiency \cite{TunableQF}, spin triplet ground states \cite{SpinDefects,Review:QuantumSensing}, shelving states \cite{AllOpticalControl}, as well as a large non-linear excitation cross-section \cite{NonLinearExcitation}. This hints at a rich underlying level structure of the plethora of possible atomic defects within the large bandgap host material.
Understanding the atomic structure of these defects will shed light on their optical levels and the underlying symmetry. If known precisely, a rich level structure is a great resource for quantum technologies. 
Hence, knowledge of the defects' atomic structure is of uttermost importance for using such quantum emitters in a scalable way and for the development of novel quantum sensing schemes \cite{Review:QuantumSensing}.
A large portion of emitters in hBN investigated so far exhibit ZPL emission in the yellow spectral region around $575$\,nm ($2.16$\,eV). Recent efforts have connected this emission to carbon-related atomic origins  \cite{IdentifyingCarbon,CarbonVacancy,CarbonAndVacancy,CarbonTrimers,CarbonAnnealing,DefectFormation}. Optical characteristics can be studied and compared with ab-initio calculations for different carbon defects to further narrow down the exact origin. To this end, photoluminescence excitation (PLE) as well as stimulated emission depletion experiments have been conducted on emitters in the yellow region \cite{AbsorptionEmissionMechanism,LowTempElectronPhononInteraction,STEDSpectroscopy,ResonantandPhononAssistedCoherentControl,CarbonVacancy}.
Following these previous studies, we perform photoluminescence excitation spectroscopy on a large number of emitters in this spectral region. In our measurements of a large set of emitters, we find a correlation in the excitation characteristics by a value of around $165$\,meV, which we interpret as a preferred coupling to a single distinct phonon mode. These findings, in combination with ab-initio predicitions, help us to identify the most likely atomic origin out of the carbon-related defects proposed in \cite{FPIdentification,LowTempElectronPhononInteraction,AbInitioIdentification,FPCalculation,CarbonTrimers}.

%%%%%%%%%%%%%%%%%%%%%%%%%%%%%%%%%%%%%%%%
\begin{figure*}[ht]
\centering
\includegraphics[width=\textwidth]{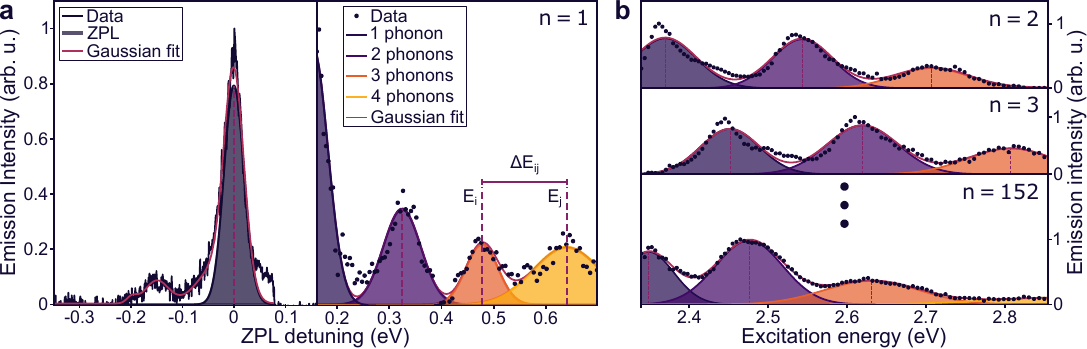}
\caption{Exemplary photoluminescence emission and excitation spectra. a) Side by side depiction of the fluorescence emission (left) and excitation (right) spectrum of a single emitter, with the respective energies (x-axis) given as the detuning from the ZPL. Energetic differences $\Delta E_{ij}$ are extracted from the positions of the fitted transitions $E_i$, $E_j$. The emission spectrum is fitted by multiple Gaussian components (see supplementary 3), from which the zero phonon line is extracted. b) Examplary PLE spectra from a total of 152 evaluated sets of data within the emission range from $555$\,nm to $585$\,nm as shown in figure~\ref{fig:3}~a.}
\label{fig:2}
\end{figure*}
%%%%%%%%%%%%%%%%%%%%%%%%%%%%%%%%%%%%%%%%
\section{Methods}
In order to characterize the optical properties of single photon emitters in hBN, we study commercially available hBN nanoflakes on silicon substrate in a homebuilt confocal fluorescence microscope. The samples undergo a consecutive annealing step at $1000$°C under a constant nitrogen flow for a duration of one hour. Single emitters are excited by a pulsed ($80$\,MHz) supercontinuum source with an accousto optical tunable filter. The excitation laser light is then filtered out and the fluorescence signal from the sample is collected in a single-mode fiber. The emitted light is then either directed to a spectrograph or two avalanche single photon detectors in a Hanbury-Brown-Twiss configuration. For a more detailed description of the setup and sample preparation, the reader is referred to supplementary~1.
From confocal scans with an excitation wavelength of $520$\,nm at a mean excitation power of $300$\,µW fluorescent spots are selected automatically and characterized (see supplementary~2). The excitation power is chosen due to the typically observed saturation power around this value under pulsed excitation at $520$\,nm (compare figure~1~a). On the selected spots, we first record photoluminescence excitation spectra by sweeping the excitation wavelength in steps of $1$\,nm from $430$\,nm to $530$\,nm at a fixed power of $50$\,µW, which is the maximum constant average power our laser system could deliver over this wavelength range. The emission count rate and second-order autocorrelation function are then measured for a duration of $100$\,s at an excitation wavelength of $520$\,nm at $300$\,µW excitation power. From these measurements we analyze the emission stability and confirm the single-photon nature of the emission. Furthermore, we measure the saturation and the photoluminescence (PL) spectrum of the emitter at $520$\,nm excitation. Exemplary measurements for a selected emitter are shown in figure~\ref{fig:1}~a.
%%%%%%%%%%%%%%%%%%%%%%%%%%%%%%%%%%%%%%%%
\begin{figure}[!ht]
\centering
\includegraphics[width=\columnwidth]{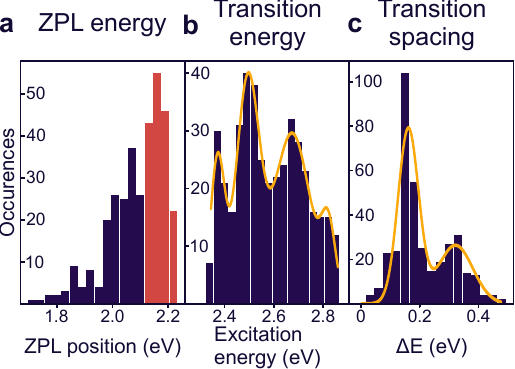}
\caption{Overview of the measured optical features of the PL and PLE spectra. The orange lines (b and c) show Gaussian fits to the densities of the respective properties.
a) Encountered zero phonon lines across all measured emission spectra. The red colored part represents the range between $2.115$\,eV and  $2.232$\,eV, on which the further evaluation was carried out.
b) Extracted positions of transitions as retrieved from the Gaussian fits of the excitation spectra (see figure~\ref{fig:2}).
c) Occurrences of energy differences between transitions in the excitation spectra of individual emitters.}
\label{fig:3}
\end{figure}
%%%%%%%%%%%%%%%%%%%%%%%%%%%%%%%%%%%%%%%%
Using this procedure a total of 6143 fluorescent spots have been selected, out of which 3079 showed signs of photobleaching and were discarded during the measurement sequence. The main limiting factor for the further analysis procedure is the blinking behavior of the majority of remaining fluorescent spots. This behavior makes the distinction between significant features of the fluorescent characteristics and temporal instabilities challenging. Thus, only 364 emitters were selected from the remaining 3064 fluorescent spots based on sufficient temporal emission stability as well as single photon purity. The selection criteria are the same for all emitters and are evaluated automatically to avoid any bias in the data (see supplementary~2).
On the remaining set of suitable emitters peaks in the recorded PLE spectrum are identified and fitted by a multi-Gaussian function (see supplement~3). In this way information about the transition distribution of individual emitters is retrieved. The recorded PL spectra are analyzed similarly for their zero phonon line and phonon contributions  \cite{PhononEmissionAbsorption,NewInsights} (see supplementary~3).
From the fits of the PLE spectra the positions and overall spacing, i.e., energetic distances between distinct transitions of individual emitters are derived. Exemplary evaluation of the measured PLE spectra is shown in figure~\ref{fig:2}. Consistent spacings between transitions in the PLE spectra can be explained by coupling to a distinct phonon mode in the well-known Huang-Rhys model (see figure~\ref{fig:1}~c), which phenomenologically describes the interaction between electrons and distinct phonon modes within the surrounding lattice \cite{HuangRhys,HR-applications}. By comparing these couplings with the theoretically predicted partial Huang-Rhys factors from ab-initio calculations, the atomic origin of the hosted defects can be narrowed down.

\section{Results}
The distribution of measured ZPLs (figure~\ref{fig:3}~a) shows that the majority of the observed emitters exhibit zero phonon line emission around $575$\,nm. Further peaks in the ZPL distribution are not significant. As recent studies suggest \cite{IdentifyingCarbon,FPIdentification,CarbonAndVacancy} this predominant yellow emission of single photon emitters in hBN is tied to carbon-related defects. Out of all the different possibilities three distinct candidates, namely $C_2C_N$, $C_2C_B$, and $V_NC_B$ (see figure~\ref{fig:1}~b), show good agreement between ab-initio calculations and experiments \cite{LowTempElectronPhononInteraction,LowTempElectronPhononInteraction,ResonantandPhononAssistedCoherentControl,STEDSpectroscopy}. In order to further fingerprint these carbon defects, we restrict the analysis to emitters with ZPL around $575$\,nm ($2.15$\,eV). This way, a set of 152 emitters remains, as highlighted in figure~\ref{fig:3}~a. Across all measured transition energies we find four local maxima (fig. \ref{fig:3}~b) in the corresponding density. The outer maxima can suffer from errors since we discard peaks in the PLE spectra close to the edge of our experimental range. We therefor focus on the distances between the found transition energies (see  fig.~\ref{fig:2}~a). The density of all extracted energy differences $\Delta E$ between transitions of individual emitters shows two distinct peaks around $158$\,meV and $317$\,meV (fig.~\ref{fig:3}~c).\newline
%%%%%%%%%%%%%%%%%%%%%%%%%%%%%%%%%%%%%%%%
\begin{figure*}[t]
\centering
\includegraphics[width=\textwidth]{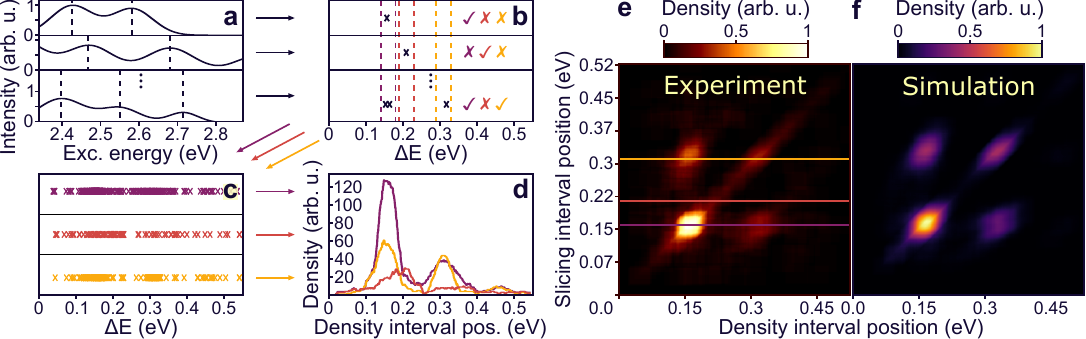}
\caption{a) -- d) Selection procedure to build subsets of emitters based on occuring energy differences in their PLE spectra. a) and b) From the fits of their PLE spectra transitions and their spacings are exctrated for all emitters. Emitters are selected, if the energy difference between two of their transitions falls within a variable interval ($50$\,meV width). c) The transition spacings for all emitters within these variable subsets. d) The density of these differences is evaluated among all conditional subsets.
\ \newline
e) Correlations among the measured features in the photoluminescence excitation spectra.
The x-axis are the densities of occuring energetic separations on conditional subsets. The center of the slicing interval for building these subsets as described in a) -- d) is depicted on the y-axis. f) shows the same evaluation of on a simulated data set with predominant spacing of $165$\,meV between transitions.}
\label{fig:4}
\end{figure*}
%%%%%%%%%%%%%%%%%%%%%%%%%%%%%%%%%%%%%%%%
To test if the higher-order maximum can be interpreted as multiple excitations of a single phonon mode, we check whether or not these extracted local maxima are correlated. We do this by performing the previous evaluation on non-disjoint subsets of emitters. These subsets are constructed by chosing a variable energy interval and selecting emitters that exhibit at least one energy difference between their transitions within this interval (fig.~\ref{fig:4}~a and b). The density of occurring energy differences is then evaluated on each subset (fig.~\ref{fig:4}~c and d). The evaluation over all subsets is depicted as a heatmap (fig.~\ref{fig:4}~e), with the centers of the density intervals and the chosen energy intervals given on the x- and y-axis, respectively. The isolated lobes of elavated intensity show that the local maxima at $158$\,meV and $318$\,meV are indeed correlated. This characteristic feature is indicative of a phonon-assisted excitation process by coupling to harmonic oscillators of a fixed fundamental energy, as depicted in the Huang-Rhys model. It should be highlighted that since this evaluation concerns differences in the PLE spectra, the second local maxima in the density plot corresponds to at least a threefold excitation of the respective fundamental phonon mode.
To determine the contributions of different phonon energies in these results, we simulate the expected behaviour of the PLE data in the presence of well-known discrete phononic contributions. This is done by creating a toy data set that matches the selected spectral region around $575$\,nm. We focus on three discrete modes located at $165$\,meV, $190$\,meV, and $100$\,meV. The first two values are chosen because of the dominant partial Huang-Rhys factors for these modes for the defects under consideration \cite{FPCalculation,CarbonTrimers,FPIdentification,LowTempElectronPhononInteraction}. The mode at $100$\,meV is selected heuristically to simulate the visibly increased density around this value (compare figure~3~c).
Starting from the ZPL, we distribute further transition lines step by step. The distance from the previous transition is selected from the above mentioned set of phonon energies at fixed relative probabilities. With this approach the PLE data of an emitter can exhibit transitions at fixed energetic spacings or a mixture of phononic modes. By varying the relative probabilities of the different components, we find that a ratio $25:2:2$ for the respective phonon energies of $165$\,meV, $190$\,meV, and $100$\,meV yields the best agreement with the data. Increasing the content of the latter two modes leads to significant deviations from the experimental data.

\section{Discussion}
From the measured PLE spectra, we find local maxima in the fluorescence distanced by multiples of $165$\,meV. From this observation, it is possible to draw conclusions about the atomic origin of these defects via their predicted phonon coupling from first principle calculations. As pointed out in several theoretical studies \cite{FPIdentification,LowTempElectronPhononInteraction,AbInitioIdentification,FPCalculation,CarbonTrimers}, two likely candidates, the $C_2C_N$ and $C_2C_B$ carbon trimers, show good agreement in terms of the theoretically proposed electron-phonon-coupling (partial Huang-Rhys factors) and experimental results. 
Out of these two possibilities, the simulated PL and PLE spectra of the $C_2C_N$ center show stronger coupling to phonon modes around $190$\,meV\footnote{The same holds true for the carbon dimer $C_BC_N$, on which the two trimers are based (see figure~\ref{fig:1}~d).}. In contrast, the $C_2C_B$ trimer shows good agreement with the observed coupling to the phonon mode around $165$\,meV. In addition, the simulations presented in \cite{LowTempElectronPhononInteraction} indicate that higher order phonon sidebands of the $C_2C_N$ defect constitute less resolved traces of the two competing modes, whereas they appear less ambiguous for the $C_2C_B$ center. The observed correlations involving the third order optical phonon mode are thus more likely to stem from the $C_2C_B$ center.
Another defect that could potentially match the observed emission around $575$\,nm, is the $C_BV_N$ carbon-vacancy center. This defect, however, exhibits larger partial Huang-Rhys factors for low-energy phonons \cite{LowTempElectronPhononInteraction,FPInvestigation,LowTempElectronPhononInteraction}, making it less likely to be the origin behind the observed emission properties.
Our method therefor serves as an identification tool via the exclusion principle. Out of four proposed defects with emission energies in the yellow region, the $C_2C_B$ shows the best agreement with the observed emission characteristics in this study.
Another supporting factor for the $C_2C_B$ center as the origin is the additional annealing step in a nitrogen-enriched atmosphere. The higher relative nitrogen concentration during the annealing favors the formation of the $C_2C_B$ over the $C_2C_N$ center due to its lower formation energy according to \cite{CarbonTrimers}. 

Due to the statistical nature of our study, this does not rule out that some emitters may still be related to other defects.
In order to put the experimental data into an objective framework it is important to highlight the limitations of this work. Temperature influences the resolution of the PLE spectra via thermal broadening \cite{LowTempElectronPhononInteraction,TempDependence,DonorAcceptor}. Similarly mixing of the two  competing modes in the higher order phonon excitations could lead to larger errors. These effects, however, will in most cases lead to the identification of the more dominant mode, which favours defect identification via its partial Huang-Rhys factors.

\section{Conclusion}
Across a set of 152 measured emitters in the spectral emission range between $2.115$\,meV and $2.232$\,meV we found a predominant coupling to the $E_{2g}$ phonon mode of the surrounding lattice. In particular, a large portion of emitters exhibit emission upon excitation via three optical phonons.
This observation can be used to connect the exact atomic configuration of the majority of single photon emitters in this range to the $C_2C_B$ carbon trimer (compare figure~\ref{fig:1}~b) based on theoretical predictions. The identification of the atomic origin of the quantum emission in hBN is a pivotal step toward the implementation of hBN in quantum applications such as quantum computing \cite{Review:QuantumComputing1,TwoPhotonInterference}, quantum key distribution \cite{QKD1Edited,QKD2,Review:HOM}, or quantum sensing \cite{Review:QuantumSensing,Review:QuantumSensing2,StrainSensing}. In order to make emitters in hBN a useful resource, detailed knowledge of its level system is a key requirement. Additionally, the exact identification of the spectral emission and excitation characteristics is paramount for the integration of these emitters into photonic structures. With the study presented here, a decisive step into this direction has been taken.
To follow this promising lead further emitters could be chosen even more restrictively. A smaller range of the ZPL position or consideration of the spectral density of the PL spectrum, lifetime, and photon yield could increase the certainty of the defect identification. Furthermore, the experiments presented in \cite{STEDSpectroscopy} could be extended to larger detunings. This way the interaction between the optical dipole and electrical field distortions induced by higher order phonon modes could be taken into account.

\section{Acknowledgements}
This work was funded by Project Nos. EMPIR 20FUN05 SEQUME and EMPIR 20IND05 QADeT. These projects have received funding from the EMPIR programme co-financed by the Participating States and from the European Union’s 2020 research and innovation programme. This work was also funded by the Deutsche Forschungsgemeinschaft (DFG, German Research Foundation) under Germany’s Excellence Strategy within the Cluster of Excellence QuantumFrontiers (EXC 2123, Project No. 390837967) and within the Cluster of Excellence PhoenixD (EXC 2122, Project No. 390833453).

\section{Author declaration}
\subsection{Conflicts of interest}
The authors have no conflicts of interest to disclose.

\subsection{Author contributions}
P. Tieben performed the experiment, evaluated the results, and wrote the manuscript. A. W. Schell designed and supervised the study. All authors discussed and interpreted the data.

\section{Data availability}
All raw data used in this study, i.e., all 6143 individual fluorescent spots characterized, are available here: https://doi.org/10.25835/q8oa042o.

\end{singlespace}

\clearpage
\bibliography{ms_bib}

\end{document}

\title{Supplementary Information: Fingerprinting Defects in hBN via Multi-Phonon Excitation}
\author
{Pablo Tieben,$^{1,2\ast}$ Andreas W. Schell,$^{1,2,3}$\\
$^{1}$Physikalisch-Technische Bundesanstalt, Bundesallee 100, 38116 Braunschweig, Germany\\
$^{2}$Institute for Solid State Physics, Gottfried Wilhelm Leibniz University, Appelstrasse 2, 30167 Hannover, Germany\\
$^{3}$Institute of Semiconductor and Solid State Physics, Johannes Kepler University,\\
Altenberger Strasse 69, 4040 Linz, Austria\\
$^\ast$E-mail:  tieben@qute.uni-hannover.de
}

\maketitle

\begin{singlespace}

\section{Experimental setup and sample preparation}
\subsection{Experimental Setup}
In order to characterize the emission properties of single photon emitters (SPEs) in hexagonaal boron nitride (hBN), we use a homebuilt confocal microscope setup. The emitters are excited via pulsed excitation by a supercontinuum white light laser source (NKT; SuperK Fianium-FIU15) with a spectral selection of the excitation wavelength by an accousto-optic tunable filter (NKT; SuperK SELECT). The excitation laser is reflected by a longpass dichroic mirror (Semrock; Beamsplitter HC 552) and is focussed on the sample by an air objective (Olympus; MPLAPON) with numerical aperture of $0.95$. To facilitate scanning, the sample is mounted on a nano positioner (PiezoSystemJena; Tritor 100). The residual laser light is filtered from the signal by a $550$\,nm longpass (Thorlabs; FELH550) and a $800$\,nm shortpass (Thorlabs; FESH800) interference filter before it is coupled into a single mode fiber (Thorlabs; P1-630A). The signal is then split by a fiber beam splitter (Thorlabs; TM50R3F2A) and detected in a Hanbury-Brown-Twiss setup consisting of two fiber-coupled single photon avalanche detectors (Laser Components; COUNT-100C- FC). For spectral analysis the signal is directed to a spectrograph (Princeton Instruments; SpectraPro HRS500) equipped with a CCD camera (Princeton Instruments; PIXIS: 100B). All spectra are recorded with a grating constant of $150$\,lines/mm.

\subsection{Sample preparation}
Our samples consist of commercially available hBN nanoflakes dispersed in a mixture of ethanol and water ($50:50$). The flakes exhibit lateral dimensions of $50$--$200$ nanometers and consist of $1$--$5$ monolayers \cite{hBN-flakes}. The solution is dropcast onto silicon wafers in five consecutive steps ($20$\,µl volume per step). To allow for a more controlled deposition of the flakes, the substrate is heated up to about $85$ degrees Celsius on a hot plate and the solution is left to dry completely after each step. After the deposition, the samples undergo a consecutive annealing step for one hour at a  temperature of $1000$ degrees Celsius under a constant nitrogen flow of $0.64$\,l h\textsuperscript{-1} at a pressure of $4.2$\,mbar.

\subsection{Nano-flake characterization}
%%%%%%%%%%%%%%%%%%%%%%%%%%%%%%%%%%%%%%%%%%%%%%%%%%%%%%%%%%%%%%%%%%%%
\begin{figure}[!ht]
\centering
\includegraphics[width=\columnwidth]{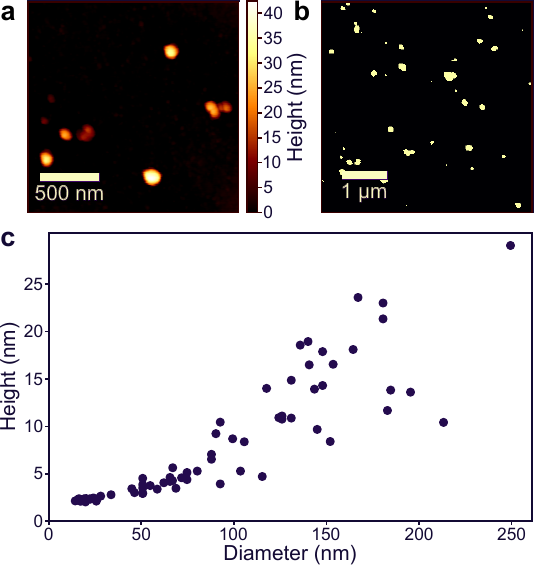}
\caption{AFM images of the samples as prepared in \textbf{section 1 B}. a) Tilt and background-corrected AFM image of an $1.8$~x~$1.8$\,µm\textsuperscript{2} area. b) Highlighted flakes from an AFM image of an $4.7$~x~$4.7$\,µm\textsuperscript{2} area for the analysis. c) Scatter plot of the measured spatial dimensions of hBN nano-flakes.}
\label{fig:1}
\end{figure}
%%%%%%%%%%%%%%%%%%%%%%%%%%%%%%%%%%%%%%%%%%%%%%%%%%%%%%%%%%%%%%%%%%%%
In order to determine the dimensions and average number of layers of the hBN nano-flakes we measure their lateral size and height in an atomic force microsope (AFM) (Core AFM; Nanosurf). The measurements are performed in dynamic force mode in order to preserve the sample integrity for consecutive measurements. Figure~\ref{fig:1}~a shows an examplary scan of a $1.8$~x~$1.8$\,µm\textsuperscript{2} area. Before the flake analysis the scans are corrected for the sample tilt by two linear fits along the x- and y-axis.
In the following we automatically identify flakes by selecting pixels of a mimimal height of $2$\,nm. Connected pixels are identified and counted as a single flake. The height of individual flakes is then retrieved as the average height over all selected pixels. Figure~\ref{fig:1}~b shows the selected flakes from an $4.7$~x~$4.7$\,µm\textsuperscript{2} scan. Out of 73 analyzed nano-flakes, we find a minimum height of $2.03$\,nm and a maximum height of $29.77$\,nm. With a lattice constant of $3.33$\,Angstrom along the axis perpendicular to the hBN layer structure these values correspond to a number of layers of $6.1$ and $89.4$ respectively. The average height of all measured emitters is $7.61$\,nm with a standard deviation of $6.37$\,nm, indicating a broad distribution of layer numbers. The average number of layers is thus found to be $22.85$ with a standard deviation of $19.13$ layers.\\
For analysis of the lateral dimensions we extract the surface area of the selected flakes. For simplicity we assume circular flakes to determine the lateral size as the corresponding diameter. By this approach we find a minimal and maximal diameter of $14.03$\,nm and $249.46$\,nm, respectively. The average diameter of all 54 flakes is $84.59$\,nm with a standard deviation of $58.98$\,nm. All measured height values and corresponding lateral dimensions can be found in figure~
\ref{fig:1}~c. Since a lot of flakes lie directly above the threshold of $2$\,nm, it is reasonable to assume that the actual distribution would be shifted towards smaller heights. For a deeper analysis of the quality of the crystal lattice in therms of the atomic composition we refer the reader to \cite{ProlongedStabiliy}.

%%%%%%%%%%%%%%%%%%%%%%%%%%%%%%%%%%%%%%%%%%%%%%%%%%%%%%%%%%%%%%%%%%%%
\begin{figure*}[!ht]
\centering
\includegraphics[width=\textwidth]{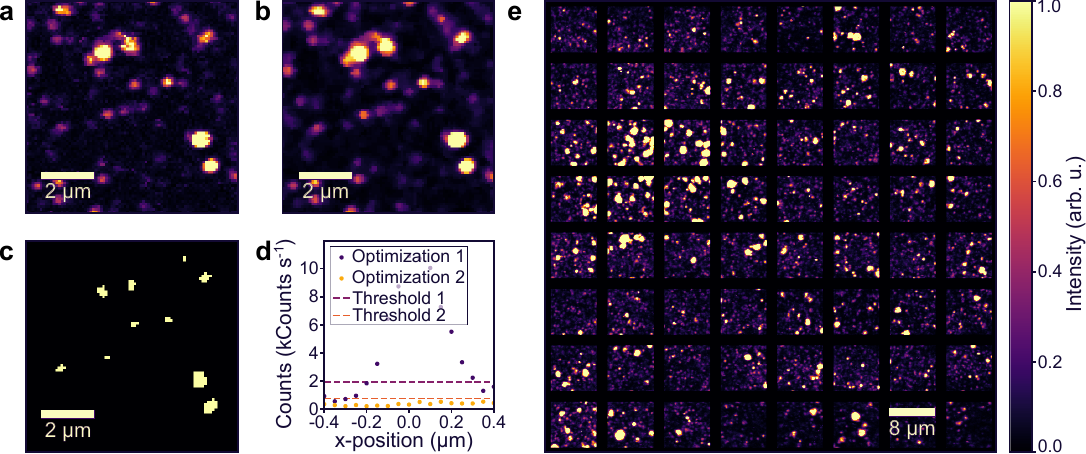}
\caption{Overview of the automatic selection process of fluorescent spots. a) Confocal scan of an $8$\,x\,$8$\,µm\textsuperscript{2} area under $520$\,nm excitation at $300$\,µW excitation power. b) Scan filtered by a Gaussian filter with $3$\,x\,$3$ kernel size. c) Selected emitters from the scan based on their brightness. Pixels that are selected possess values of one, while the background is set to zero. d) Comparison between the x-optimization process before (blue) and after (orange) a  measurement indicating photobleaching in between. The dashed lines depict the respective threshold value, which have to be surpassed in order to continue the measurements. e) Full range ($80$\,x\,$80$\,µm\textsuperscript{2}) of the nanopositioner covered by $8$\,x\,$8$\,µm\textsuperscript{2} scans with $2$\,µm of space between adjacent scans. All fluorescence scans are scaled up in intensity to provide better visibility.}
\label{fig:2}
\end{figure*}
%%%%%%%%%%%%%%%%%%%%%%%%%%%%%%%%%%%%%%%%%%%%%%%%%%%%%%%%%%%%%%%%%%%%
\section{Automatic measurement sequence and data selection}
\subsection{Emitter selection and measurements}
We select emitters automatically from confocal scans of $8$\,$x$\,$8$\,µm\textsuperscript{2} with a step size of $0.1$\,µm by comparing the brightness of each pixel with its surrounding pixels. The resulting scans are smoothened by a Gaussian filter with a kernel size of $3$\,x\,$3$ pixels\footnote{The filter function ``$median\_filter(input,size=3)$'' is taken from the python package ``scipy.ndimage''.}. From these filtered scans pixels that exceed the brightness of the sixth neighbouring pixel in each direction by at least a factor of four are selected. Connected pixels are then grouped together and interpreted as one emitter\footnote{To connect the pixels the function ``$label(input, np.ones((3, 3)))$'' is taken from the python package ``scipy.ndimage''.}. 
After the identification process, the positioner is set to the coordinates of the selected emitters, which is followed by an xyz-optimization step to amount for setup drift. Then the photoluminescence excitation spectrum (PLE), second-order autocorrelation function and count rate, photoluminescence emission spectrum (PL), and saturation measurements are performed. The PLE spectrum is measured for an integration time of $500$\,ms per wavelength with a step size of the excitation wavelength of $1$\,nm and a constant excitation power of $50$\,µW. This value is chosen due to the limited output power of the laser source across the whole wavelength range from $430$\,nm to $530$\,nm. Count rate, autocorrelation, emission spectrum, and saturation measurements are all performed under $520$\,nm excitation. The count rate (binwidth=$10$\,ms) and autocorrelation (binwidth=$200$\,ps) are measured simultaneously under constant mean excitation power of $300$\,µW (typical saturation power) for a duration of $100$\,s. The PL spectrum is recorded with $300$\,µW excitation power and is integrated for $40$\,s. The saturation of the emitters is measured in $25$ separate values distributed logarithmically up to a maximum excitation power of $500$\,µW integrated over $500$\,ms per step. Each measurement is followed by an additional xyz-optimization and checked for possible photobleaching. Further measurements are aborted in the case of a bleaching event in order to save measurement time. Bleaching of the emitter is assumed if the maximum is not at least $3.5$ times higher than the minimum brightness for the x- and y-optimization, respectively (see figure~\ref{fig:2}~d). Bleaching was detected on a total of 3079 out of 6143 measured emitters. This procedure is repeated by covering the full scan range of the nanopositioner ($80$\,x\,$80$\,µm\textsuperscript{2}) with smaller scans while leaving $2$\,µm of space in between to avoid doubles in the data (see figure~\ref{fig:2}~e).

%%%%%%%%%%%%%%%%%%%%%%%%%%%%%%%%%%%%%%%%%%%%%%%%%%%%%%%%%%%%%%%%%%%%
\begin{figure*}[t]

\centering
\includegraphics[width=\textwidth]{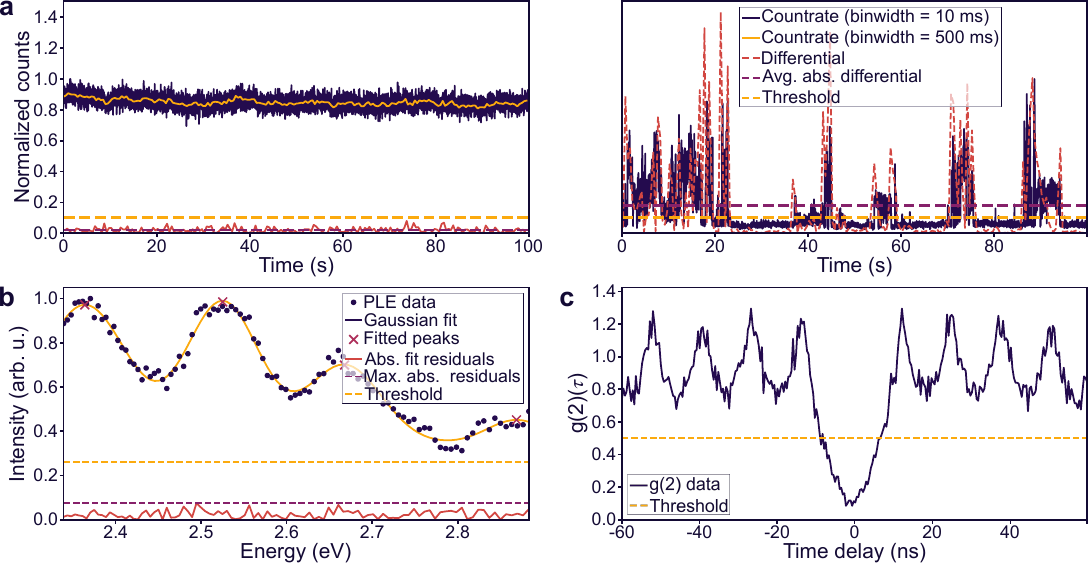}
\caption{Selection criteria for the recorded data sets. a) Comparison between a stable (left) and unstable (right) emission count rate on a $10$\,ms (blue) and $500$\,ms (yellow) binning window. The dashed red line shows the derivative on the $500$\,ms scale. Negative values are omitted to increase visibility. The purple and yellow horizontal dashed lines show the average absolute value of the derivative and the threshold for the data selection, respectively. b) Multi-Gaussian fit function to the recorded PLE data together with the absolute residuals. The horizontal purple line shows the maximum of the resiudals and the horizontal yellow line indicates the threshold for the selection. c) Second-order autocorrelation measurement exhibiting clear antibunching. The dashed yellow line shows the treshold value of $0.5$}
\label{fig:3}

\end{figure*}
%%%%%%%%%%%%%%%%%%%%%%%%%%%%%%%%%%%%%%%%%%%%%%%%%%%%%%%%%%%%%%%%%%%%
\subsection{Data selection}
Another phenomenon, namely blinking (instabilities in the emission flux of single emitters), may hinder the reliable extraction of their properties. In order to separate the physical features, i.e., peaks in the PLE spectrum, from these instabilities, the following steps are implemented to automatically select single emitters.
\begin{itemize}
\item[1)]{The measured count rate is evaluated for average brightness and emission stability. The average count rate must exceed a minimum value of $8000$\,counts/s. Furthermore the average absolute of the differential of the normalized\footnote{The normalization is performed with respect to the maximum value, i.e., all values are divided by the maximum recorded value. This holds for this and all following measurements, unless otherwise specified.} count rate must be below a value of $0.1$ (see figure~\ref{fig:3}~a) on average in order for the emitter to be selected for further evaluation. This evaluation is performed on a binning window of $500$\,ms (binning window of the PLE measurements) to ensure consistency.}
\item[2)]{The recorded autocorrelation measurement must exhibit a value of $g^{(2)}(0)<0.5$. This criterium ensures that the majority of detected photons indeed stem from a single emitter.}
\item[3)]{The multi-Gaussian fit (see supplementary~3 for details) to the PLE spectra must show a decent agreement with the data. To ensure this, the residuals between fit and normalized data are used. If the maximum absolute value of the retrieved residuals lies below $0.26$ (see figure~\ref{fig:3}~c), the data set of the respective emitter is selected.}
\end{itemize}

%%%%%%%%%%%%%%%%%%%%%%%%%%%%%%%%%%%%%%%%%%%%%%%%%%%%%%%%%%%%%%%%%%%%
\begin{figure*}[!ht]

\centering
\includegraphics[width=\textwidth]{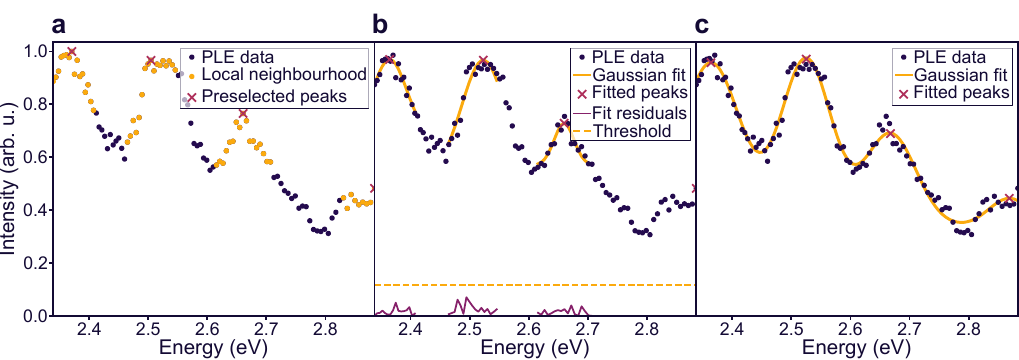}
\caption{Outline of the peak finding and fitting routine. a) Preselection of peak values based on their intensity in comparison to a neighborhood of eight data points in each direction. b) Gaussian fits around the preselected peaks and their absolute residuals. The dashed purple line shows the threshold value for the residuals in order to be considered for further evaluation. c) Multi-Gaussian fits to the data and extracted peak points.}
\label{fig:4}

\end{figure*}
%%%%%%%%%%%%%%%%%%%%%%%%%%%%%%%%%%%%%%%%%%%%%%%%%%%%%%%%%%%%%%%%%%%%
\section{Fitting of the PLE and PL spectra}
\subsection{Peak finder and Multi-Gaussian fit}
In order to fit the recorded PLE and PL spectra, we pre select peaks in the normalized data by the following approach: A data point is identified as a peak value if its value is larger than all of the surrounding x values in each direction, where x takes a value of $8$ and $25$ data points for the PLE and PL data respectively (see figure~\ref{fig:4}~a). On the respective neighbourhood, a Gaussian is then fitted and the absolute values of the residuals as well as the height of the Gaussian are taken as the final criterium. If the maximum of the absolute of the residuals is below $0.12$ ($0.15$) and the peak height is larger than $0.1$ ($0.06$), the data point is considered a peak in the PLE (PL) spectrum (see figure~\ref{fig:4}~b). 
Peaks at the edges of the data are selected simply by the first criterium and used for the Gaussian fit in the following to allow for a good fit over the full experimental range. The position of a peak at the edge of the data, however, cannot be identified with confidence and these points are thus discarded for further evaluation.
%%%%%%%%%%%%%%%%%%%%%%%%%%%%%%%%%%%%%%%%%%%%%%%%%%%%%%%%%%%%%%%%%%%%
\begin{figure}[!ht]

\centering
\includegraphics[width=\columnwidth]{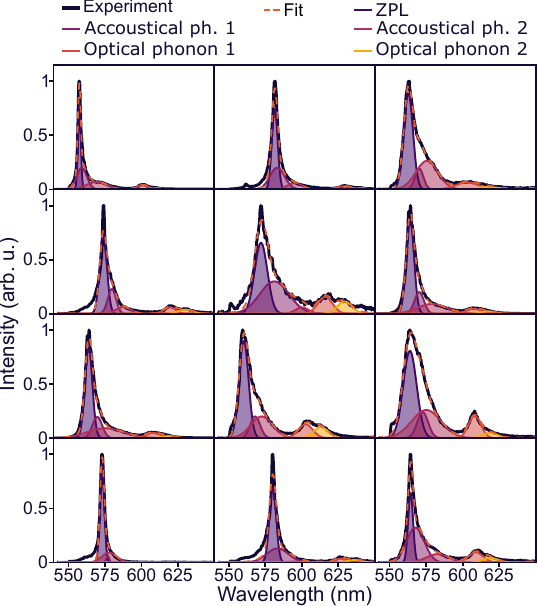}
\caption{Exemplary emission spectra of different emitters. All specta are recorded under $520$\,nm excitation at $300$\,µW and are integrated over 40 s. The full spectrum is fitted by multiple Gaussian contributions consisting of the zero phon line, two low-energy accoustical phonons and two optical phonons.}
\label{fig:5}

\end{figure}
%%%%%%%%%%%%%%%%%%%%%%%%%%%%%%%%%%%%%%%%%%%%%%%%%%%%%%%%%%%%%%%%%%%%
%%%%%%%%%%%%%%%%%%%%%%%%%%%%%%%%%%%%%%%%%%%%%%%%%%%%%%%%%%%%%%%%%%%%
\begin{figure*}[!ht]
\centering
\includegraphics[width=\textwidth]{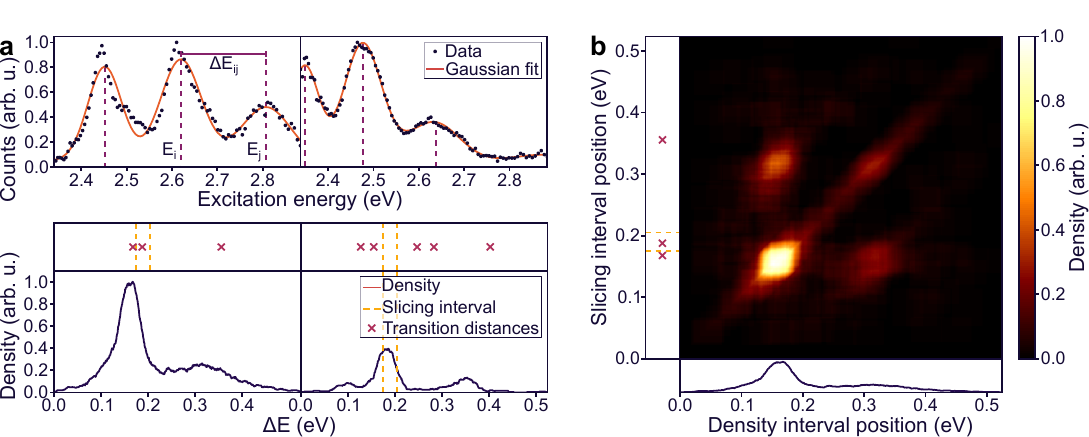}
\caption{Evaluation process on conditional subsets. a) Creation of a sub data set based on a chosen energy interval. If at least one extracted energy difference from the PLE data (top) lie within the interval (middle), the PLE data is selected for the subset. The left and right panels show exemplary PLE spectra with transition spacings $\Delta E_{ij}$, that are selected and discarded for the chosen subset, respectively. The density of energetic spacings in the PLE data is evaluated on the resulting subsets. The bottom panels show the density of these values on the full (left) and sliced (right) data set based on the shown interval. b) Colormap of the full density evaluation based on varying the slicing interval over the full experiment range. To the left and to the bottom of the image are visual representations of the density given on the x-axis and the slicing interval varied over the y-axis.}
\label{fig:6}
\end{figure*}
%%%%%%%%%%%%%%%%%%%%%%%%%%%%%%%%%%%%%%%%%%%%%%%%%%%%%%%%%%%%%%%%%%%%
The preselected peaks are used to fit a multi Gaussian function to the spectra. For the fitting procedure the ``curve\_fit'' function from the ``scipy.optimize'' python package is used with a least square method. The initial guessed function is estimated by the peak positions, heights, and widths (according to the distances to the surrounding minima\footnote{The minima are selected analogously to the peaks from the inverted data.}) from the previous step.

\subsection{Photoluminescence emission spectra}
According to \cite{PhononEmissionAbsorption} the emission spectrum of most SPEs in hBN is composed of several parts. Most importantly, the ZPL is accompanied by a phonon sideband (or several). Furthermore, the ZPL is usually asymmetrically broadened, which can be explained by coupling to low-energy phonon modes (see figure~\ref{fig:5}). Due to this fact, the ZPL position is extracted simply as the maximum of the recorded PL spectrum for practical reasons. This alternative approach only influences the accuracy of the ZPL assignment on a negligible scale (compare figure~\ref{fig:5}).
Across the set of all measured emitters we find vastly varying phonon contributions in the photoluminescence emission spectra (Debye-Waller factor). For all emission spectra exhibiting notable phonon contributions we find a visible asymmetry in the first order optical phonon sideband, which can be fitted by two optical phonon modes around $165$\,meV and $190$\,meV, respectively.

\section{Correlated density plot}
\subsection{Evaluation}
To investigate correlations in the measured density of spacings between transitions, we start by evaluating the density on conditional subsets. Subsets are built based on a chosen interval and individual emitters are selected if at least one of the differences between their transitions lies within this interval. Figure~\ref{fig:6}~a shows the measured PLE spectra (top) of two exemplary emitters and the extracted energetic differences (bottom) between their transitions. The first one satisfies the selection condition given by the highlighted interval, while the second one is discarded from the respective subset. The density of occurring energy differences can be evaluated and compared against the full data set (see figure~\ref{fig:6}~a (bottom)). Finally, we vary the position of the chosen slicing interval and perform the density evaluation on each resulting subset. The final plot (see figure~\ref{fig:6}~b) shows this evaluation as a heatmap, with the density intervals (interval width = $50$\,meV, compare figure~\ref{fig:6}~a) given on the x- and the position of the slicing intervals (interval width = $40$\,meV,) given on the y-axis (compare figure~\ref{fig:6}~a). The evaluation of the distances from the measured ZPLs is performed analogously over the respective values.

%%%%%%%%%%%%%%%%%%%%%%%%%%%%%%%%%%%%%%%%%%%%%%%%%%%%%%%%%%%%%%%%%%%%
\begin{figure}[!t]
\centering
\includegraphics[width=\columnwidth]{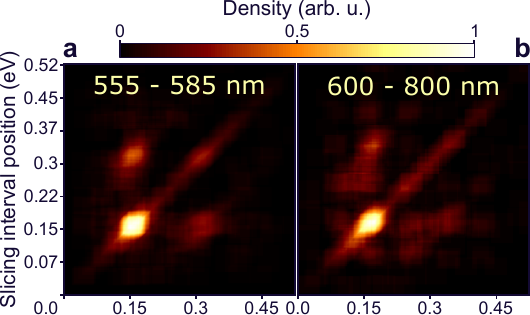}
\caption{Comparison between emitters from different regions of zero phonon line emission. a) Colormap of the full density evaluation based on varying the slicing interval for emitters with zero phonon line position between $555$ and $585$\,nm. b) The same evaluation for emitters with zero phonon lines between $600$ and $800$\,nm.}
\label{fig:7}
\end{figure}
%%%%%%%%%%%%%%%%%%%%%%%%%%%%%%%%%%%%%%%%%%%%%%%%%%%%%%%%%%%%%%%%%%%%
\section{Comparison between different emission ranges}
For completion we compare the previously introduced analysis for emitters from different spectral emission regions. Figure~\ref{fig:7}~a shows the chosen emitter range from the main text with ZPL wavelengths between $555$ and $585$\,nm. Emitters with ZPL wavelengths in the range of $600$ and $800$\,nm are shown in figure~\ref{fig:7}~b. The sets contain $152$ and $114$ emitters, respectively. 
One notable difference is the slight blue shift of the main optical phonon mode for the second set of emitters. In addition, correlations by different values can be observed, visible by the repeating patterns in several lines in fig.~\ref{fig:7}~b.
However, these features should be considered with care as there are several factors compromising the reliability of the analysis. First, the detuning from the ZPL is larger and therefor the phonon-assisted excitation is expected to be less efficient. Secondly, mixing of phonon energies becomes more impactful at higher orders. This makes the identification of local maxima in the PLE spectra less reliable, as the signal to noise ratio is less favourable. 
Furthermore, the distrubution of different ZPL is sparser in the range from $600$ to $800$\,nm (compare figure 3 a) in the main text). A meaningful approach for comparison with ab-initio calculations on a well-chosen smaller interval is thus more difficult at higher wavelengths.

\subsection{Data simulation}
\begin{figure}[!t]
\centering
\includegraphics[width=\columnwidth]{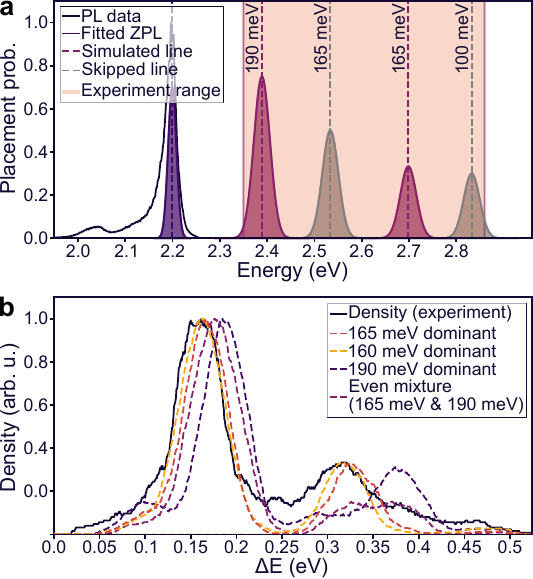}
\caption{Simulation and evaluation of the photoluminescence data. a) Distribution of transition lines based on the ZPL (purple) value inside of the experiment range (orange shaded area). The position is chosen from a set of values with respective probabilities. The height and width of the lines represent the descending placement probability with increasing number of placed transitions and the Gaussian jitter of $17$\,meV, respectively. Skipped lines are depicted in gray. b) Density of energy differences between transition lines for simulated data sets with different contributions of the three chosen modes. The set with strongly dominating mode at $160$\,meV (yellow) yields the best match with the data set (blue), while the mode at $190$\,meV does not match the pattern in the data.
}
\label{fig:8}
\end{figure}

To simulate a data set of emitters with arbitrarily spaced transitions, we start with the measured ZPL positions within the range from $555$\,nm	 and $585$\,nm, as in the main text. To increase the statistical significance the set of measured ZPL positions is then duplicated sevenfold, resulting in a set of 1064 emitters while preserving the original ZPL distribution. 
For every emitter, we now start to place toy transitions consecutively at distances from a discrete set of values and respective probabilities from the ZPL within the range of the experiment between $2.34$\,eV and $2.88$\,eV (see figure~\ref{fig:8}~a). The distance is chosen independently on each step and the transition is then placed at this distance from the last line (starting from the ZPL), allowing for a mixture of occurring distances (modes) for single emitters (see figure~\ref{fig:8}~a).
To simulate noise in the data, we introduce a Gaussian jitter to the placement procedure (see figure~5~a). Lastly, the decreasing coupling strength with increasing number of involved phonons, i.e., the (on average) decreased peak height for transitions further from the ZPL, has to be considered in the simulation. This is implemented via a coin event to decide whether or not a transition is placed or not with successively decreasing probability for each placed transition. If a  transition is not placed the placement procedure is continued from the respective position of the skipped data point, which can result in empty spots in an otherwise homogeneous pattern (compare figure~\ref{fig:8}~a). 
We focus on a set of three discrete modes located around $165$\,meV, $190$\,meV, and $100$\,meV. The first two values are picked in accordance to the theoretically predicted dominant partial Huang-Rhys factors for the $C_2C_B$ and $C_2C_N$ centers, respectively (\cite{FPCalculation,CarbonTrimers,FPIdentification,LowTempElectronPhononInteraction}). The mode at $100$\,meV is chosen to simulate the visibly increased density around this value (compare figure~3~d in the main text). The relative probabilities for these modes and the Gaussian jitter have been heuristically derived as $\{25,2,2\}$ and $17$\,meV, respectively ,to give the best match to the experimental data. Similarly, the probability function to determine the placement has been derived as
\begin{equation}
p(m,n)=\frac{3}{m+n+1}.
\end{equation}
%%%%%%%%%%%%%%%%%%%%%%%%%%%%%%%%%%%%%%%%%%%%%%%%%%%%%%%%%%%%%%%%%%%%
%%%%%%%%%%%%%%%%%%%%%%%%%%%%%%%%%%%%%%%%%%%%%%%%%%%%%%%%%%%%%%%%%%%%
Here $n$ is twice the number of lines that have been generated but not necessarily placed within the range of the experiment. This simulates the descending overall intensity of most observed PLE spectra and thus the decreased probability to identify transitions further from the ZPL. $m$ is an integer starting at one that is incremented each time a line is skipped and reset to one when a line is placed. The motivation for this is the observation, that most inspected PLE spectra indicate a missing line in the middle of the excitation range to be less likely than a homogeneously distributed pattern with decreasing intensity (compare figure~\ref{fig:6}~a and figure~3~a in the main text). 
We find overall good agreement between experiment and simulation for a predominant mode at $165$\,meV, however with a slight shift of the experimental data towards lower energetic correlations in comparison to the simulated data (see figure~\ref{fig:8}~b). A lower energetic mode at $160$\,meV (compare \cite{FPIdentification}) yields a better fit than the otherwise suggested value of $165$\,meV. This shift in phonon energy might arise from the elevated temperature in this study, as compared to other experimental studies and most other ab-initio calculations concerning cryogenic temperatures.

\end{singlespace}

\clearpage
\bibliography{supp_bib}